\documentclass[Royal,sagev,times, doublespace]{sagej1}
\usepackage{epstopdf}
\usepackage{moreverb,url}
\usepackage[colorlinks,bookmarksopen,bookmarksnumbered,citecolor=red,urlcolor=red]{hyperref}
\usepackage[utf8]{inputenc} 
\newcommand\BibTeX{{\rmfamily B\kern-.05em \textsc{i\kern-.025em b}\kern-.08em
T\kern-.1667em\lower.7ex\hbox{E}\kern-.125emX}}

\begin{document}

\runninghead{Nyaga \textit{et al.}}

\title{ANOVA model for network meta-analysis of diagnostic test accuracy data}

\author{Victoria Nyaga\affilnum{1,2}, Marc Aerts\affilnum{2} and Marc Arbyn\affilnum{1}}

\affiliation{\affilnum{1}Scientific Institute of Public Health, Unit of Cancer Epidemiology/Belgian Cancer Center, Belgium\\
\affilnum{2}Hasselt University, Center for Statistics, Belgium}

\corrauth{Victoria Nyaga
Scientific Institute of Public Health, Unit of Cancer Epidemiology/Belgian Cancer Center,
Juliette Wytsmanstreet 14,
1050, Brussels,
Belgium.}

\email{victoria.nyaga@uhasselt.be}

\begin{abstract}
Network meta-analysis (NMA) allow combining efficacy information from multiple comparisons from trials assessing different therapeutic interventions for a given disease and to estimate unobserved comparisons from a network of observed comparisons. Applying NMA on diagnostic accuracy studies is a statistical challenge given the inherent correlation of sensitivity and specificity.

A conceptually simple and novel hierarchical arm-based (AB) model which expresses the logit transformed sensitivity and specificity as sum of fixed effects for test, correlated study-effects and a random error associated with various tests evaluated in given study is proposed. We apply the model to previously published meta-analyses assessing the accuracy of diverse cytological and molecular tests used to triage women with minor cervical lesions to detect cervical precancer and the results compared with those from the contrast-based (CB) model which expresses the linear predictor as a contrast to a comparator test.

The proposed AB model is more appealing than the CB model in that it yields the marginal means which are easily interpreted and makes use of all available data and easily accommodates more general variance-covariance matrix structures.

\end{abstract}


\maketitle

\section{Introduction}
Network meta-analyses (NMA) have classically been used to extend conventional pairwise meta-analyses by combining and summarizing direct and indirect evidence on multiple `therapeutic' interventions for a given condition when the set of evaluated interventions/treatments differs among studies. By borrowing strength from the indirect evidence, there is a potential gain in  precision of the estimates \cite{Higgins}. Furthermore, the estimates may be less biased and more robust. Such an approach uses the data efficiently and is line with the principle of intention-to-treat (ITT) \cite{Fisher} in randomized clinical trials which requires that all valid available data should be used even when a part of the data is missing.

In a diagnostic test accuracy study, an index test and possibly one or more comparator tests are administered to each tested subject. A standard or reference test or procedure is also applied to all the patients to classify them as having the target condition or not. The patients results are then categorized by the index and reference test as true positive, false positive, true negative and false negative. The diagnostic accuracy of the index test is represented as a bivariate outcome and is typically expressed as sensitivity and specificity at a defined test cutoff. Differences due to chance, design, conduct, patients/participants, interventions, tests and reference test imply there will be heterogeneity often in opposite direction for the two typical accuracy outcomes: sensitivity and specificity. While traditional meta-analyses allow for comparison between two tests, there are often multiple tests for the diagnosis of a particular disease outcome. To present the overall picture, inference about all the tests for the same condition and patient characteristics is therefore required. The simultaneous analysis of the variability in the accuracy of multiple tests within and between studies may be approached through a network meta-analysis.

In combining univariate summaries from studies where the set of tests differs among studies two types of linear mixed models have been proposed. The majority of network meta-analyses express treatment effects in each study as a contrast relative to a baseline treatment in the respective study \cite{Higgins, Lumley}. This is the so called contrast-based (CB) model. Inspired by the CB models developed for interventional studies, Menten and Lesaffre (2015) \cite{Menten} introduced a CB model for diagnostic test accuracy data to estimate the average log odds ratio for sensitivity and specificity of the index test relative to a baseline or comparator test. 

The second type of models is the classical two-way ANOVA model with random effects for study and fixed effect for tests \cite{Senn, Whitehead, Piepho12}, the so called arm-based (AB) model.  The AB model is based on the assumption that the missing arms or tests are missing at random. While the two types of models yield similar results for the contrasts with restricted maximum likelihood (REML) procedures, the CB model is generally not invariant to changes in the baseline test in a subset of studies and yields an odds ratio (OR) making it difficult to recover information on the absolute diagnostic accuracy (the marginal means), relative sensitivity or specificity of a test compared to another or differences in accuracy between tests, measures that are easily interpretable and often used in clinical epidemiology. It is common knowledge that the OR is only a good approximation of relative sensitivity/specificity when the outcome is rare but this is often not the case in diagnostic studies. Moreover, the AB model is simpler when the baseline/comparator treatment varies from one study to another or when the number of tests varies substantially among studies. By accommodating more complex variance-covariance structures AB models have been shown to be superior to CB models \cite{Zhang}.

We apply the two-way ANOVA model in a diagnostic data setting by extending the AB model in two ways: 1. using two independent binomial distributions to describe the distribution of true positives and true negatives among the diseased and the healthy individuals, 2. inducing a correlation between sensitivity and specificity by introducing correlated and shared study effects. The resulting generalized linear mixed model is analogous to randomized trials with complete block designs or repeated measures in analysis of variance models where studies are equivalent to blocks. The main assumption is that,  results missing for some tests and studies are missing at random. This approach is efficient because the correlation structure allows the model to borrow information from the `imputed' missing data to obtain adjusted sensitivity and specificity estimates for all the tests.

\section{Motivating dataset}
To illustrate the use of the proposed model in network meta-analysis of diagnostic test accuracy data, we analyse data on a diversity of cytological or molecular tests to triage women with equivocal or mildly abnormal cervical cells  \cite{Arbyn12, Arbyn13a, Arbyn13b, Roelens, Verdoodt}. A Pap smear is a screening test used to detect cervical precancer. When abnormalities in the Pap smear are not high grade, a triage test is needed to identify the women who need referral for further diagnostic work-up. There are several triage options, such as repetition of the Pap smear or HPV DNA or RNA assays.  HPV is the virus causing cervical cancer  \cite{Bosch}. Several other markers can be used for triage as well, such as p16 or the combinations of p16/Ki67 which are protein markers indicative for a transforming HPV infection~\cite{Roelens, Arbyn09} .

The data are derived from a comprehensive series of meta-analyses on the accuracy of triage with HPV assays, cervical cytology or molecular markers applied on cervical specimens in women with minor cervical abnormalities \cite{Arbyn12, Arbyn13a, Arbyn13b, Roelens, Verdoodt}. Two patient groups with minor cytological abnormalities were distinguished: women with ASC-US (atypical squamous cells of unspecified significance) and LSIL (low-grade squamous intraepithelial lesions). Studies were included in the analysis if they performed besides one or more triage test a verification with a reference standard based on colposcopy and biopsies. 

In total, the accuracy of 11 tests for detecting cervical precancer were evaluated. Labelled 1 to 11 the tests were: hrHPV DNA testing with HC2 (HC2), Conventional Cytology (CC), Liquid-Based Cytology (LBC), generic PCRs targeting hrHPV DNA (PCR) and commercially available PCR-based hrHPV DNA assays such as: Abbott RT PCR hrHPV, Linear Array, and Cobas-4800; assays detecting mRNA transcripts of five (HPV Proofer) or fourteen (APTIMA) HPV types HPV types; and protein markers identified by cytoimmunochestry such as: p16 and p16/Ki67, which are over-expressed as a consequence of HPV infection. Two levels of precancer (disease) were considered: intraepithelial neoplasia lesion of grade two or worse (CIN2+) or of grade three or worse (CIN3+). 125 studies with at least one test and maximum of six tests were included allowing assessment of the accuracy of the eleven triage tests. In 

\begin{figure}[h]
\fbox{\includegraphics[width=\textwidth, height=\textheight,keepaspectratio]{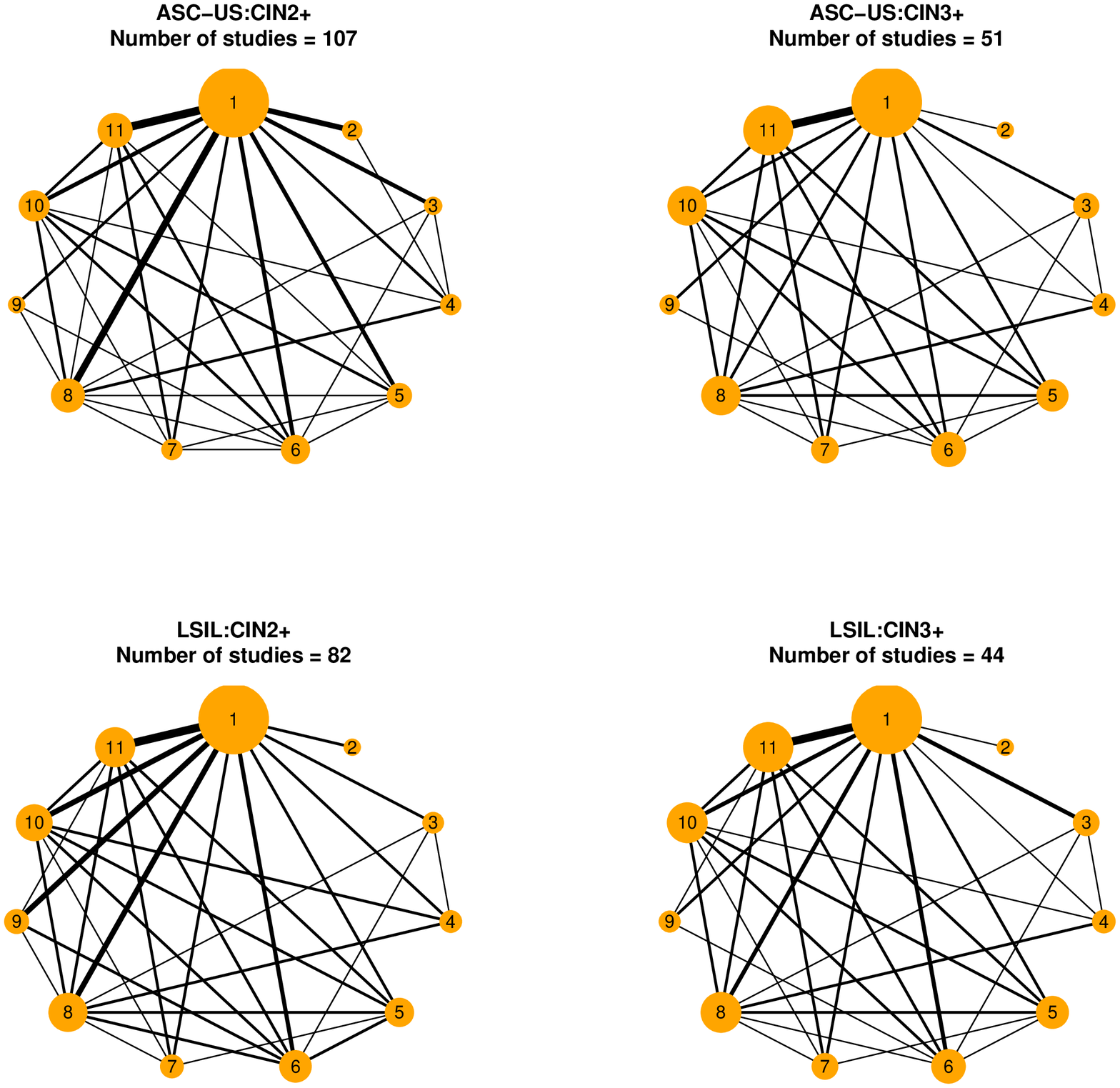}}
\caption{Network plot of all included tests\protect\endnotemark[1] by triage\protect\endnotemark[2] group (women with ASC-US or LSIL cytology) and the outcome\protect\endnotemark[3] (CIN2+ or CIN3+).\label{Fig:1}}	
\end{figure}
The size of the nodes in figure~\ref{Fig:1} is proportional to the number of studies evaluating a test and thickness of the lines between the nodes is proportional to the number of direct comparisons between tests. The size of the node and the amount of information in a node consequently influence the standard errors of the marginal means and the relative measures. From the network plot, test 1 (HC2) and test 11 (APTIMA) were the most commonly assessed tests. The network in figure~\ref{Fig:1} is connected.   
\section{Methodology}
Suppose there are \textit{K} tests and \textit{I} studies. Studies assessing two tests ($ k = 2 $) are called `two-arm' studies while those with $k  > 2 $ are ‘multi-arm’ studies. For a certain study \textit{i}, let $(Y_{i1k}, ~Y_{i2k})$ denote the true positives and true negatives,  $(N_{i1k}, ~N_{i2k})$ the diseased and healthy individuals and $(\pi_{i1k}, ~\pi_{i2k})$ the `unobserved'  sensitivity and specificity respectively with test \textit{k} in study \textit{i}. Given study-specific sensitivity and specificity, two independent binomial distributions describe the distribution of true positives and true negatives among the diseased and the healthy individuals as follows;
\begin{equation}\label{Eq:0}
Y_{ijk} ~|~ \pi_{ijk}, ~x_i ~\sim~ bin(\pi_{ijk}, ~N_{ijk}), ~i ~=~ 1, ~\ldots I, ~j ~=~ 1, ~2, ~k~ = 1,~ \ldots ~K,
\end{equation}				
where $x_i$ generically denotes one or more covariates, possibly affecting $\pi_{ijk}$. In the next section, we present the recently introduced contrast-based model~\cite{Menten} followed by our proposed arm-based model to estimate the mean as well as comparative measures of sensitivity and specificity. 
\subsection{Contrast-based model}
By taking diagnostic test $T_K$ as the baseline, Menten and Lessafre (2015) \cite{Menten} proposed the following model,
\begin{align}\label{Eq:1}
	logit(\pi_{ijk}) ~=~ \theta_{ijk} \nonumber\\
	\theta_{ij1} ~=~ \mu_{ij} ~+~ (K ~-~ 1)\times \frac{\delta_{ij1}}{K} ~-~ \frac{\delta_{ij2}}{K} ~-~ \frac{\delta_{ij3}}{K} ~-~ \ldots ~\frac{\delta_{ij(k-1)}}{K} \nonumber\\
	\theta_{ij2} ~=~ \mu_{ij} ~-~  \frac{\delta_{ij1}}{K} ~+~ (K ~-~ 1)\times\frac{\delta_{ij2}}{K} ~-~ \frac{\delta_{ij3}}{K} ~-~ \ldots ~\frac{\delta_{ij(k-1)}}{K} \nonumber\\
	\theta_{ijK} ~=~ \mu_{ij} ~-~  \frac{\delta_{ij1}}{K} ~-~ \frac{\delta_{ij2}}{K} ~-~ \frac{\delta_{ij3}}{K} ~-~ \ldots ~\frac{\delta_{ij(k-1)}}{K} \nonumber\\
	with \nonumber\\
	(\delta_{i11}, ~\delta_{i12}, ~\delta_{i21}, ~\delta_{i22}, ~\ldots, ~\delta_{i1(K-1)}, ~\delta_{i2(K-1)}) ~\sim~ N(\boldsymbol{\nu}_\delta, ~\boldsymbol{\Sigma}) \nonumber\\
	and \nonumber \\
	\nu_\delta ~=~ (\nu_{\delta11}, ~\nu_{\delta21}, ~\nu_{\delta12}, ~\nu_{\delta22} ~\ldots, ~\nu_{\delta1(K-1)}, ~\nu_{\delta2(K-1)}) 
\end{align} 
The $\boldsymbol{\nu}_\delta$ represents the average log odds ratio for sensitivity and specificity of the \textit{K~-~1} tests compared to the baseline test $T_K$. There are known difficulties in estimating the variance-covariance matrix $\boldsymbol{\Sigma}$ since each sampled matrix should be positive-definite \cite{Daniels}. The authors therefore recommend a diagonal or block diagonal variance-covariance matrix $\boldsymbol{\Sigma}$. While this reduces model complexity and difficulty in estimation, such a covariance matrix accounts for the correlation between contrasts but ignores correlation between sensitivity and specificity. Moreover, the model identification becomes difficult as the number of tests included increases.
 
The authors estimate the absolute accuracy of the tests from the estimated  $logit^{-1}( \mu_{jk})$ as follows
\begin{align}\label{Eq:2}
\mu_{j1} ~=~ logit^{-1}(E(\mu_j)) ~+
&~ \frac{K-1}{K}\times \nu_{\delta{j1}} ~-~ \frac{1}{K}\times\nu_{\delta{j2}} ~-~ \frac{1}{K}\times\nu_{\delta{j3}} ~-~ \ldots ~-~ \frac{1}{K}\times\nu_{\delta{j(K-1)}} \nonumber\\
\mu_{j2} ~=~ logit^{-1}(E(\mu_j)) ~-
&~ \frac{1}{K}\times\nu_{\delta{j1}} ~+~ \frac{K-1}{K}\times\nu_{\delta{j2}} ~-~ \frac{1}{K}\times\nu_{\delta{j3}} ~-~ \ldots ~-~ \frac{1}{K}\times\nu_{\delta{j(K-1)}} \nonumber\\
\mu_{j3} ~=~ logit^{-1}(E(\mu_j)) ~-
&~ \frac{1}{K}\times\nu_{\delta{j1}} ~-~ \frac{1}{K}\times\nu_{\delta{j2}} ~+~ \frac{K-1}{K}\times\nu_{\delta{j3}} ~-~ \ldots ~-~ \frac{1}{K}\times\nu_{\delta{j(K-1)}} \nonumber\\
\mu_{jK} ~=~ logit^{-1}(E(\mu_j)) ~-
&~ \frac{1}{K}\times\nu_{\delta{j1}} ~-~ \frac{1}{K}\times\nu_{\delta{j2}} ~-~ \frac{1}{K}\times\nu_{\delta{j3}} ~-~ \ldots ~-~ \frac{1}{K}\times\nu_{\delta{j(K-1)}} 	
\end{align}
where $logit^{-1}(E(\mu_j))$ is the average probability of testing positive/negative. Equation~\ref{Eq:2} estimates the accuracy of tests for a hypothetical study with random-effects equal to zero but not the meta-analytic estimates as will be explained in the next section.
\subsection{Arm-based model}
Consider a design where there is at least one test per study. The study serves as a block where all diagnostic accuracy tests are hypothetically evaluated of which some are missing. This modelling approach has potential gain in precision by borrowing strength from studies with single tests as well as multi-arm studies. The proposed single-factor design with repeated measures model is written as follows
\begin{align}\label{Eq:3}
logit(\pi_{ijk}) = \mu_{jk} + \eta_{ij} + \delta_{ijk} \nonumber\\
\begin{pmatrix}
\eta_{i1} \nonumber\\
\eta_{i2} 
\end{pmatrix} \sim N \bigg (\begin{pmatrix}
								0 \nonumber\\
								0 
								\end{pmatrix}, \boldsymbol{\Sigma} \bigg ) \nonumber\\
\boldsymbol{\Sigma} = \begin{bmatrix}
\sigma^2_1 ~~~~ \rho\sigma_1\sigma_2 \nonumber\\
\rho\sigma_1\sigma_2 ~~~~ \sigma^2_2 
\end{bmatrix} \nonumber\\
(\delta_{ij1}, \delta_{ij1}, \ldots \delta_{ijK}) \sim N(\textbf{0}, diag(\tau^2_j))
\end{align}
where $\mu_{1k}$ and $\mu_{2k}$ are the mean sensitivity and specificity in a hypothetical study with random-effects equal to zero respectively. $\eta_{ij}$ is the study effect for healthy individuals \textit{(j = 1)} or diseased individuals \textit{(j = 2)} and represents the deviation of a particular study \textit{i} from the mean sensitivity (j=1) or specificity (j=2),  inducing between-study correlation. The study effects are assumed to be a random sample from a population of such effects. The between-study variability of sensitivity and specificity and the correlation thereof is captured by the parameters $\sigma_1^2$, $\sigma_2^2$, and $\rho$ respectively. $\delta_{ijk}$ is the error associated with the sensitivity (\textit{j=1}) or specificity (\textit{j=2}) of test \textit{k} in the $i^{th}$ study. Conditional on study \textit{i}, the repeated measurements are independent with variance constant across studies such that $\boldsymbol{\tau}_j^2$ = $(\tau_{j1}^2, \ldots, \tau_{jK}^2)$ is a \textit{K} dimensional vector of homogeneous variances. 

In case $\tau_{jk}^2 = \tau_j^2$ (variances homogenous across tests), the shared random element $\eta_{ij}$ within study \textit{i} induce a non-negative correlation between any two test results \textit{k} and $k\prime$ from healthy individuals \textit{(j = 1)} or from diseased individuals \textit{(j = 2)} equal to $\rho_j = \frac{\sigma_j^2}{\sigma_j^2 + \tau_j^2}$ (implying that a covariance matrix with compound symmetry). While it might seem logical to expect and allow for similar correlation between any two sensitivities or specificities in a given study, the variances $\tau_{jk}^2$ of different sensitivities or specificities of the same study may be different. In such instances, the unstructured covariance matrix is more appropriate as it allows varying variances between the tests (in which case $\boldsymbol{\tau}_j^2$ is a \textit{K} dimension vector of the unequal variances). The correlation between the $k^{th}$ and $k^{'th}$ test result is then equal to $\rho_{jkk}' = \frac{\sigma_j^2}{\sqrt{\sigma_j^2 + \tau_{jk}^2~\times (\sigma_j^2 + \tau_{jk'}^2 )}}$. $\rho_j$ or $\rho_{jkk}'$ is called the intra-study correlation coefficient which also measures the proportion of the variability in $logit(\pi_{ijk})$ that is accounted for by the between study variability. It takes the value 0 when $\sigma_j^2 = 0$ (if study effects convey no information) and values close to 1 when $\sigma_j^2$ is large relative to $\tau_j^2$ and the studies are essentially all identical. When all components of $\boldsymbol{\tau}_j^2$ equal to zero, the model reduces to fitting separate bivariate random-effect meta-analysis (BRMA)~\cite{reitsma, chu} model for each test. 

In essence, the model separates the variation in the studies into two components: the within-study variation $diag(\tau_j^2)$ referring to the variation in the repeated sampling of the study results if they were replicated, and the between-study variation $\boldsymbol{\Sigma}$ referring to variation in the studies true underlying effects.

The study-level covariate information is included in the linear predictor in Equation~\ref{Eq:3} as follows
\begin{equation} \label{Eq:17}
logit(\pi_{ijk}) = \mu_{jk} + \sum_{p = 1}^{P} \theta_{pjk} X_{pi} + \eta_{ij} + \delta_{ijk}	
\end{equation}
where $\theta_{pjk}$ is the $p^{th}$ coefficients corresponding to the $X_{pi}$ covariate in a hypothetical study with random-effects equal to zero respectively.
The population-averaged or the marginal sensitivity/specificity in the intercept-only model for test \textit{k} is estimated as
\begin{align}\label{Eq:4}
E(\pi_{ijk}) ~=~ & E(logit^{-1}(\mu_{jk} ~+~ \eta_{ij} ~+~ \delta_{ijk})) \nonumber\\
 =& \int_{-\infty}^{\infty}logit^{-1}(\mu_{jk} ~+~ \eta_{ij} ~+~ \delta_{ijk})~f(\eta_{ij})~f(\delta_{ijk})~d\eta_{ij}~d\delta_{ijk}.
\end{align}
The relative sensitivity and specificity and other relative measures of test \textit{k} (relative to test $k\prime, k \ne k\prime$) are then estimated from the marginal sensitivity or specificity of test \textit{k} and \textit{k’}.

In most practical situations, the mean structure is of primary interest and not the covariance structure. Nonetheless, appropriate covariance modelling is critical in the interpretation of the random variation in the data as well as obtaining valid model-based inference for the mean structure. Compound symmetry assumes homogeneity of variance and covariance and such restriction could invalidate inference for the mean structure when the assumed covariance structure is misspecified \cite{Altham}. When the primary objective of the analysis is on estimating the marginal means of sensitivity and specificity, the choice between compound symmetry and unstructured covariance structure is not critical because the inference procedure for the marginal means are the same. Moreover, over-parameterisation of the covariance structure might lead to inefficient estimation and potentially poor assessment of standard errors of the marginal means \cite{Verbeke}. 
\subsection{Ranking of the tests}
While ranking of tests using rank probabilities and rankograms is an attractive feature of univariate NMA, it is still a challenge to rank competing diagnostic tests especially when a test does not outperform the others on both sensitivity and specificity.  
Consider the diagnostic odds ratio (DOR) \cite{Glas} which is expressed in terms of sensitivity and specificity as
\begin{equation} \label{Eq:5}
DOR_k = \frac{sensitivity_k \times specificity_k}
			 {(1 - sensitivity_k) \times (1 - specificity_k)}.
\end{equation}
and ranges from 0 to $\infty$ with: $DOR_k > 1$ or higher indicating better discriminatory test performance, $DOR_k = 1$ indicating a test that does not discriminate between the healthy and diseased, and $DOR_k < 1$ indicating an improper test. The DOR is a single indicator combining information about sensitivity and specificity and is invariant of disease prevalence. However, the measure cannot distinguish between tests with high sensitivity but low specificity or vice-versa. 

Alternatively, the superiority of a diagnostic test could be quantified using a superiority index introduced by Deutsch et al. \cite{Deutsch} expressed as
\begin{equation} \label{Eq:6}
S_k = \frac{2a_k + c_k}
		   {2b_k + c_k},
\end{equation}
where $a_k$ is the number of tests to which test \textit{k} is superior (higher sensitivity and specificity), $b_k$ is the number of tests to which test \textit{k} is inferior (lower sensitivity and specificity), and $c_k$ the number of tests with equal performance as test \textit{k} (equal sensitivity and specificity). \textit{S} ranges from 0 to $\infty$ with; $S$ tending to $\infty $ and $S $ tending to $0$ as the number of tests to which test \textit{k} is superior and inferior increases respectively, and $S$ tending to $1$ the more the tests are equal. Since the number of tests not comparable to test \textit{k} do not enter into the calculation of \textit{S} the index for different tests may be based on different sets of tests. 
\subsection{Missing data and exchangeability }
In the models above, not all the studies provide estimates of all effects of interest because some of the components of the vector $\textbf{Y}_{ij} = (Y_{ij1}, \ldots, Y_{ijK})$ are missing. The $\textbf{Y}_{ij}$ vector can be partitioned into the observed $\textbf{Y}_{ij}^o$ and the missing $\textbf{Y}_{ij}^m$. For each component of $\textbf{Y}_{ij}$ let $\textbf{R}_{ij}$ denote a vector of missingness indicator with
\begin{equation} \label{Eq:7}
R_{ijk} = \bigg \{ \begin{matrix}
								1 ~ if ~Y_{ijk} ~is ~observed,\nonumber\\
								0 ~ otherwise. 
								\end{matrix}
\end{equation}
The joint distribution of (\textbf{Y}, \textbf{R}) given the parameters ($\beta, \phi$) is given by 
\begin{equation} \label{Eq:8}
p(y_{ij}, R_{ij}| \boldsymbol{\beta}_{j}, \boldsymbol{\phi}_{j})
\end{equation}
where
$\boldsymbol{\phi}_{j}$ contains the missingness paramaters and $\boldsymbol{\beta}_{j}$ contains $(\boldsymbol{\pi}_{ij}, ~\boldsymbol{\Sigma}, ~\rho, ~\sigma_j, ~ diag(\boldsymbol{\tau}_j))$. In a selection framework \cite{Rubin, Little} the joint distribution in Equation~\ref{Eq:8} is factorised as
\begin{equation} \label{Eq:9}
p(y_{ij}~|~ \boldsymbol{\beta}_{j}), \boldsymbol{\phi}_{j}~p(R_{ij}~|~ Y_{ij}, ~\boldsymbol{\phi}_{j}) =
p(y_{ij}^o, ~y_{ij}^m ~|~ \boldsymbol{\beta}_{j}, ~\boldsymbol{\phi}_{j})p(R_{ij}~|~ Y_{ij}^o, ~Y_{ij}^m,~ \boldsymbol{\phi}_{j})	
\end{equation}
where $p(R_{ij}~|~y_{ij}^o,~y_{ij}^m, ~\boldsymbol{\phi_{j}})$ describes the mechanism for data missingness. Assuming that the probability of missingness is conditionally independent of the unobserved data given the observed (so called missing at random (MAR)), the second part of Equation~\ref{Eq:9} simplifies to
\begin{equation} \label{Eq:10}
p(R_{ij}~|~y_{ij}^o,~y_{ij}^m,~\boldsymbol{\phi}_{j})~=~ p(R_{ij}~|~y_{ij}^o, ~ \boldsymbol{\phi}_{j}).
\end{equation}
When the parameters $\beta_{ij}$ and $\phi_{ij}$ are distinct and functionally independent, the missing data mechanism is ignorable and the Expression~\ref{Eq:10} can be dropped from the joint distribution in Equation~\ref{Eq:8}. 
Intergrating over the unknown missing values in the first part of Equation~\ref{Eq:9} yields a marginal density with the observed information which is to be evaluated
\begin{equation} \label{Eq:11}
\int_{}^{}p(y_{ij}^o, ~y_{ij}^m~|~ \boldsymbol{\beta}_{j})~dy_{ij}^m ~=~ p(y_{ij}^o~|~\boldsymbol{\beta}_{j}).
\end{equation}
Since the main objective is to be able to make valid and efficient inference about the parameters of interest and not to estimate or predict the missing data, the ignorability condition validates inference based on the observed data likelihood only. 
Conditional on $\pi_{ijk}$ the studies are assumed to be exchangeable. The observed information $Y_{ijk}$ on a given test/arm \textit{k} generically represents a point estimate of $\pi_{ijk}$ and contributes to the estimation of the fixed effects $\mu_{jk}$. 
At the second level of the hierarchy (Equation~\ref{Eq:2} and ~\ref{Eq:4}), exchangeable normal prior distributions with mean zero split the variability into between- and within-study variability. The observed data in each study contributes to the estimation of $\eta_{ij}$ while all the studies all-together contribute to the estimation of $\delta_{ijk}$ where $\delta_{ijk}$ and  $\eta_{ij}$ are considered independent samples from a population controlled by the hyper-parameters $\boldsymbol{\Sigma}$ and $\tau_j^2$ which are estimated from the observed data. The hyper-parameters also have exchangeable vague or non-informative prior distributions. 

The exchangeability assumption is applied in both the CB and the AB models but in a different manner. The CB model assumes exchangeability of tests contrasts (odds ratios) across the studies while the AB assumes exchangeability of tests effects (means) across the studies.
\subsection{Prior distributions}
We decompose the covariance matrix $\boldsymbol{\Sigma}$ into a variance and correlation matrix such that
\begin{equation} \label{Eq:12}
\boldsymbol{\Sigma} = diag(\sigma_1, \sigma_2) \times ~\boldsymbol{\Omega} \times~diag(\sigma_1, \sigma_2),
\end{equation}
where 
\begin{equation} \label{Eq:13}
\boldsymbol{\Omega} = \begin{bmatrix}
1 ~~ \rho \nonumber\\
\rho ~~ 1
\end{bmatrix}.
\end{equation}
The model is completed by specifying vague priors on the mean, variance and correlation parameters as follows
\begin{align}\label{Eq:14}
tanh^{-1}(\rho), ~\mu_{jk} \sim N(0, ~25) \nonumber\\
\tau_j, ~\sigma_j \sim U(0, ~5).
\end{align}
Since it is not clear when certain choices of prior distributions are vague and non-informative, it is necessary to vary the prior distribution and assess their influence on the parameter estimates. The following prior distributions were also used as part of a sensitivity analysis
\begin{align}\label{Eq:15}
\rho ~\sim U(-1, ~1) \nonumber\\
\tau_j, ~\sigma_j \sim cauchy(0, ~2.5).
\end{align}
An alternative prior distribution for the correlation matrix $\boldsymbol{\Omega}$ is the LKJ prior distribution with shape parameter $\nu~=~1$ or $\nu~=~2$ \cite{Lewandowski}. 
\begin{equation}\label{Eq:16}
\boldsymbol{\Omega} ~\sim ~LKJcorr(\nu) ~\propto det(\boldsymbol{\Omega})^{\nu-1} ~ for ~ \nu \ge 1,
\end{equation}
where $\nu$ controls the expected correlation with larger values favouring less correlation and vice-versa.
Other possible prior distributions for $\boldsymbol{\Sigma}$ are: the Inverse-Wishart distribution having the advantage of computational convenience but being difficult to interpret or the more relaxed scaled inverse Wishart which is a conjugate to the multivariate normal making Gibbs sampling simpler \cite{Gelman07}.
\subsection{Implementation}
The models are fitted in the Bayesian framework using Stan \cite{Stan}, a probabilistic programming language which has implemented Hamilton Monte Carlo (MHC) and No-U-Turn sampler (NUTS) \cite{Hoffman} within R 3.2.3 \cite{R} using the rstan 2.8.2 package \cite{rstan}. The Stan code for the model is provided alongside the supplementary material. We run three chains in parallel until there is convergence. Trace plots are used to visually check whether the distributions of the three simulated chains mix properly and are stationary. For each parameter, convergence is assessed by examining the potential scale reduction factor $\hat{R}$, the effective number of independent simulation draws ($n_{eff}$) and the MCMC error. It is common practice to run simulations until $\hat{R}$ is no greater than 1.1 for all the parameters. Since Markov chain simulations tend to be autocorrelated, $n_{eff}$ is usually smaller compared to the total number of draws.  To reduce autocorrelation and consequently increase $n_{eff}$, it is necessary to do thinning by keeping every $n^{th}$ (e.g. every $10^{th}$, $20^{th}$, $30^{th}$ \ldots) draw and discarding the rest of the samples. Besides, thinning saves memory especially when the total number of iterations is large.

\section{Results}
Figure~\ref{Fig:2} presents the study-specific sensitivity and specificity of all the eleven used to detect CIN2+ in ASC-US triage from all available studies and from studies that evaluated at least two tests one of them being test 1. We successively present the sensitivity and specificity of the eleven tests in triage of ASC-US and LSIL for outcomes CIN2+ and CIN3+, in figures~\ref{Fig:3}, \ref{Fig:4} and \ref{Fig:5} respectively.
Representing the pooled sensitivity and specificity, the black diamonds are estimated by the AB model from all the available studies, the red diamond by the same model but from studies with at least two tests with one of them being test 1 while the blue diamonds are estimated by the CB model from studies with at least two tests with one of them being test 1. The vertical lines represent the 95\% credible intervals. In each instance, the studies included in estimating the diagnostic accuracy estimates are in grey points underlying the diamonds.
From the study-specific grey points there was substantial variation in both sensitivity and specificity between the studies and some studies  had outlying values. It is also apparent that the number of tests evaluated differed among studies (see also supplementary material : Additional-tables.docx). 

\subsection{All available data (black diamonds)}
\subsubsection{Triage of women with ASC-US to detect CIN2+}
\begin{figure}[h] 
\fbox{\includegraphics[width=\textwidth, height=\textheight,keepaspectratio]{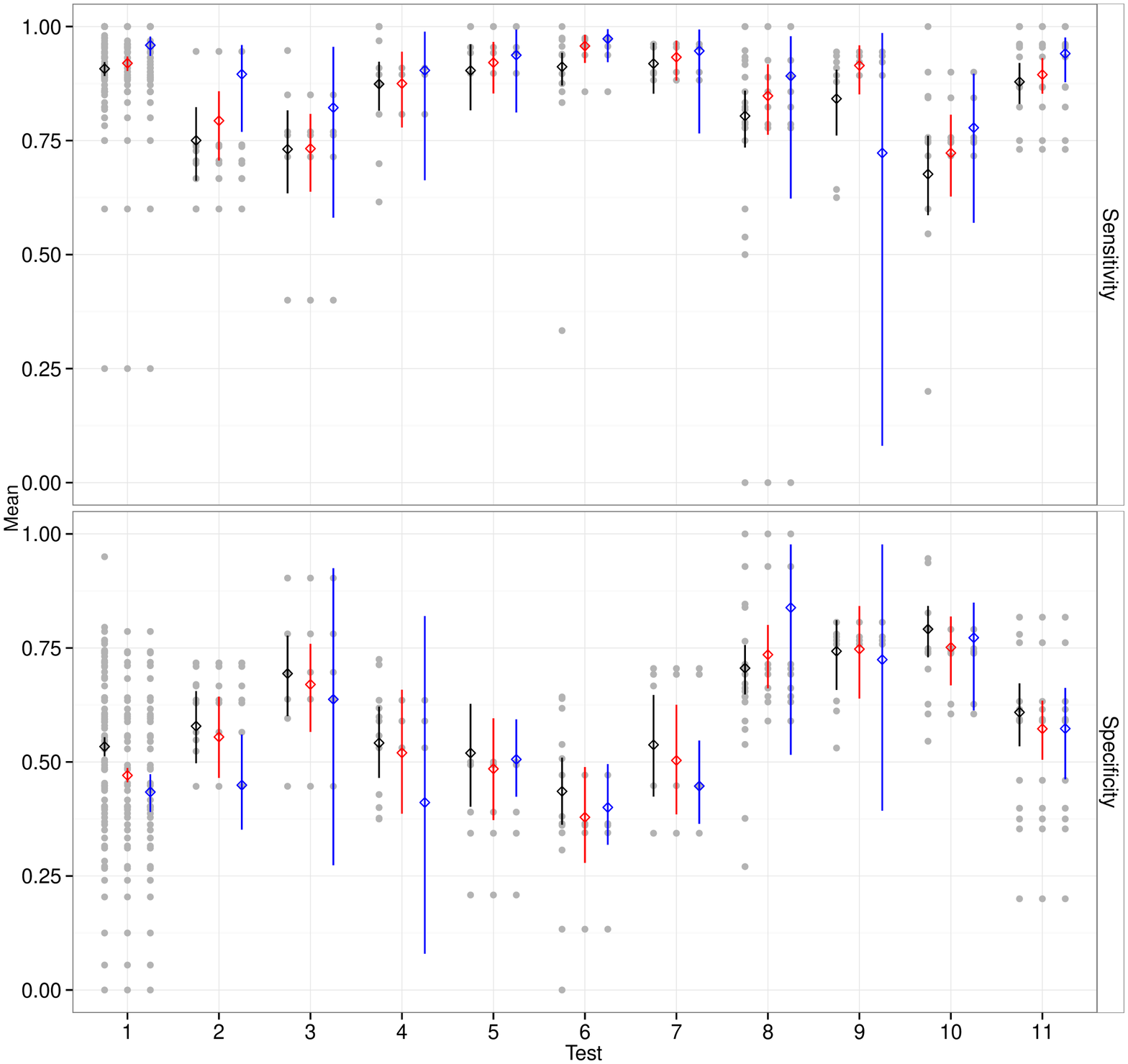}}	
 \caption{Plot of study-specific sensitivity (top) and specificity (bottom) in grey points and their corresponding pooled (diamonds) estimates with their 95\% credible intervals (vertical lines) of tests\protect\endnotemark[1] detecting CIN2+\protect\endnotemark[3] in ASC-US\protect\endnotemark[2] triage.      
 The black diamonds and vertical lines are estimated by the AB model from all the available studies, the red by the same model but from studies with at least two tests with one of them being test 1 while the blue are estimated by the CB model from studies with at least two tests with one of them being test 1.\label{Fig:2}}
  \end{figure}

According to figure~\ref{Fig:2}, test 6 (Linear Array) was the most sensitive (0.91[0.87, 0.94]) but among the least specific (0.44 [0.36, 0.51]) tests while test 10 (HPV Proofer) the least sensitive (0.68 [0.59, 0.76]) and the most specific (0.79 [0.73, 0.84]) test.  

Both the diagnostic odds ratio and superiority index in the supplementary material (Results1.xlsx) indicate that test 9 (p16/Ki67) had the best discriminatory power with a sensitivity of 0.84 [0.76, 0.91] and specificity of 0.74 [0.66, 0.81]. 
Compared to test 1 (HC2), tests 3 (LBC), 8 (p16) and 10 (HPV Proofer) were less sensitive but more specific, while tests 9 (p16/Ki67) was equally as sensitive but more specific. All other tests had similar sensitivity and specificity as test 1 (HC2) (see table~\ref{Tab:1}). 

\begin{table}[h]
	\centering
	\caption{Posterior relative sensitivity and specificity and the corresponding 95\% credible intervals of other tests relative to test 1 (HC2) in detecting CIN2+\protect\endnotemark[3] in ASC-US\protect\endnotemark[2] triage as estimated by the AB model}
	\label{Tab:1}
	\begin{tabular}{|l|l|l|l|l|l|l|l|}
	\hline
	&                             & \multicolumn{3}{l|}{Relative sensitivity} & \multicolumn{3}{l|}{Relative specificity} \\ \hline
	Label & Index test                  & Mean        & Lower        & Upper        & Mean        & Lower        & Upper        \\ \hline
	2     & Conventional Cytology (CC)  & 0.83        & 0.73         & 0.91         & 1.08        & 0.93         & 1.24         \\ \hline
	3     & Liquid-Based Cytology (LBC) & 0.81        & 0.70         & 0.90         & 1.30        & 1.12         & 1.47         \\ \hline
	4     & Non-Commercial PCR assays   & 0.96        & 0.90         & 1.02         & 1.02        & 0.87         & 1.18         \\ \hline
	5     & Abbott RT PCR hrHPV         & 1.00        & 0.90         & 1.06         & 0.97        & 0.75         & 1.18         \\ \hline
	6     & Linear Array                & 1.00        & 0.96         & 1.05         & 0.82        & 0.67         & 0.96         \\ \hline
	7     & Cobas-4800                  & 1.01        & 0.94         & 1.06         & 1.01        & 0.80         & 1.21         \\ \hline
	8     & P16                         & 0.89        & 0.81         & 0.95         & 1.32        & 1.19         & 1.44         \\ \hline
	9     & P16/Ki67                    & 0.93        & 0.84         & 1.00         & 1.39        & 1.22         & 1.54         \\ \hline
	10    & HPV Proofer(mRNA)           & 0.75        & 0.65         & 0.84         & 1.48        & 1.36         & 1.59         \\ \hline
	11    & APTIMA(mRNA)                & 0.97        & 0.91         & 1.02         & 1.14        & 1.00         & 1.26         \\ \hline
	\end{tabular}
\end{table}

\subsubsection{Triage of women with ASC-US to detect CIN3+}
\begin{figure}[h] 
\fbox{\includegraphics[width=\textwidth, height=\textheight,keepaspectratio]{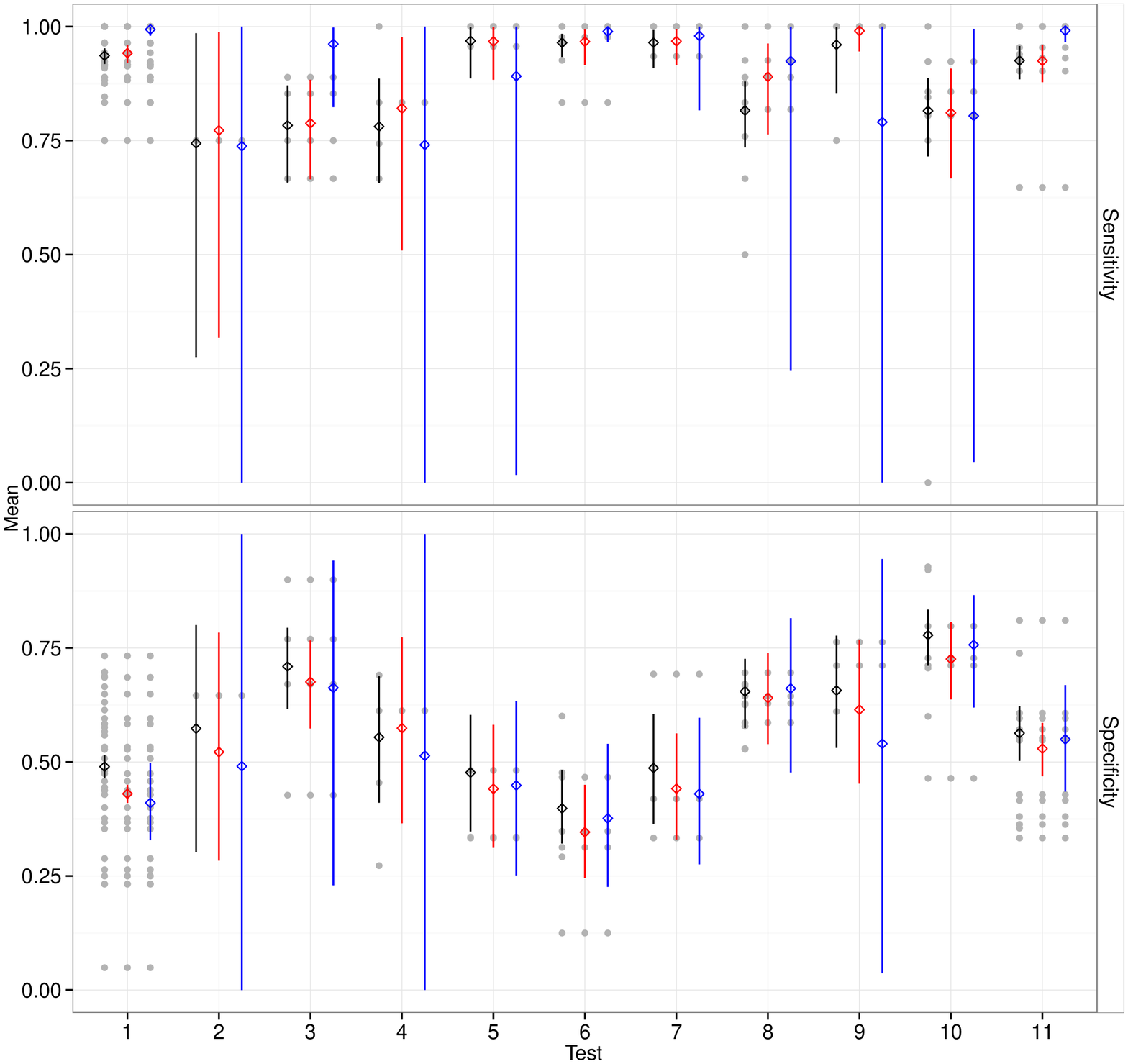}}		
\caption{Plot of study-specific sensitivity (top) and specificity (bottom) in grey points and their corresponding pooled (diamonds) estimates with their 95\% credible intervals (vertical lines) of tests\protect\endnotemark[1] detecting CIN3+\protect\endnotemark[3] in ASC-US triage\protect\endnotemark[2]. The black diamonds and vertical lines are estimated by the AB model from all the available studies, the red by the same model but from studies with at least two tests with one of them being test 1 while the blue are estimated by the CB model from studies with at least two tests with one of them being test 1.\label{Fig:3}}
 \end{figure}

It can be seen in figure~\ref{Fig:3} that test 5 (Abbott RT PCR hrHPV) was the most sensitive (0.97 [0.89, 1.00]) but among least specific (0.48 [0.35, 0.60]) tests. The diagnostic odds ratio and the superiority index (see supplementary material :Results1.xlsx) indicate that test 9 (p16/Ki67) had the best discriminatory power with sensitivity and specificity of 0.96 [0.85, 1.00] and 0.66 [0.53, 0.78] respectively.  

Relative to test 1 (HC2), test 3 (LBC), 4 (Non-commercial PCR assays), 8 (p16) and 10 (HPV Proofer) were less sensitive but more specific while tests 2 (CC), 5 (Abbott RT PCR hrHPV), 7 (Cobas-4800) were as sensitive and specific (see table~\ref{Tab:2}).

\begin{table}[h]
	\centering
	\caption{Posterior relative sensitivity and specificity and the corresponding 95\% credible interval of other tests relative to test 1 (HC2) in detecting CIN3+\protect\endnotemark[3] in ASC-US\protect\endnotemark[2] triage as estimated by the AB model}
	\label{Tab:2}
	\begin{tabular}{|l|l|l|l|l|l|l|l|}
		\hline
		&                             & \multicolumn{3}{l|}{Relative sensitivity} & \multicolumn{3}{l|}{Relative specificity} \\ \hline
		Label & Index test                  & Mean        & Lower        & Upper        & Mean        & Lower        & Upper        \\ \hline
		2  & Conventional Cytology (CC)  & 0.79 & 0.29 & 1.06 & 1.17 & 0.62 & 1.65 \\ \hline
		3  & Liquid-Based Cytology (LBC) & 0.84 & 0.71 & 0.93 & 1.45 & 1.25 & 1.63 \\ \hline
		4  & Non-Commercial PCR assays   & 0.83 & 0.70 & 0.95 & 1.13 & 0.83 & 1.43 \\ \hline
		5  & Abbott RT PCR hrHPV         & 1.03 & 0.95 & 1.08 & 0.97 & 0.70 & 1.24 \\ \hline
		6  & Linear Array                & 1.03 & 0.99 & 1.06 & 0.81 & 0.65 & 0.99 \\ \hline
		7  & Cobas-4800                  & 1.03 & 0.97 & 1.07 & 0.99 & 0.74 & 1.24 \\ \hline
		8  & P16                         & 0.87 & 0.79 & 0.94 & 1.34 & 1.14 & 1.51 \\ \hline
		9  & P16/Ki67                    & 1.03 & 0.91 & 1.08 & 1.34 & 1.08 & 1.60 \\ \hline
		10 & HPV Proofer(mRNA)           & 0.87 & 0.77 & 0.95 & 1.59 & 1.43 & 1.74 \\ \hline
		11 & APTIMA(mRNA)                & 0.99 & 0.94 & 1.03 & 1.15 & 1.02 & 1.28 \\ \hline
	\end{tabular}
\end{table}

\subsubsection{Triage of women with LSIL to detect CIN2+}
Figure~\ref{Fig:4} and the absolute diagnostic estimates presented in the supplementary material (Results1.xlsx) show that test 1 (HC2) was the most sensitive (0.94 [0.93, 0.95]) test but among the least specific (0.29 [0.27, 0.31]) tests while test 10 (HPV proofer) was the least sensitive (0.64 [0.54, 0.73]) and the most specific (0.73 [0.67, 0.78]) test detecting CIN2+ in LSIL cytology. Both the diagnostic odds ratio and superiority index presented in the supplementary material (Results1.xlsx) indicate once more that test 9 (p16/Ki67) had the best discriminatory power with an estimated sensitivity and specificity of 0.86 [0.79, 0.91] and 0.63 [0.57, 0.69]. 

\begin{figure}[h] 
\fbox{\includegraphics[width=\textwidth, height=\textheight,keepaspectratio]{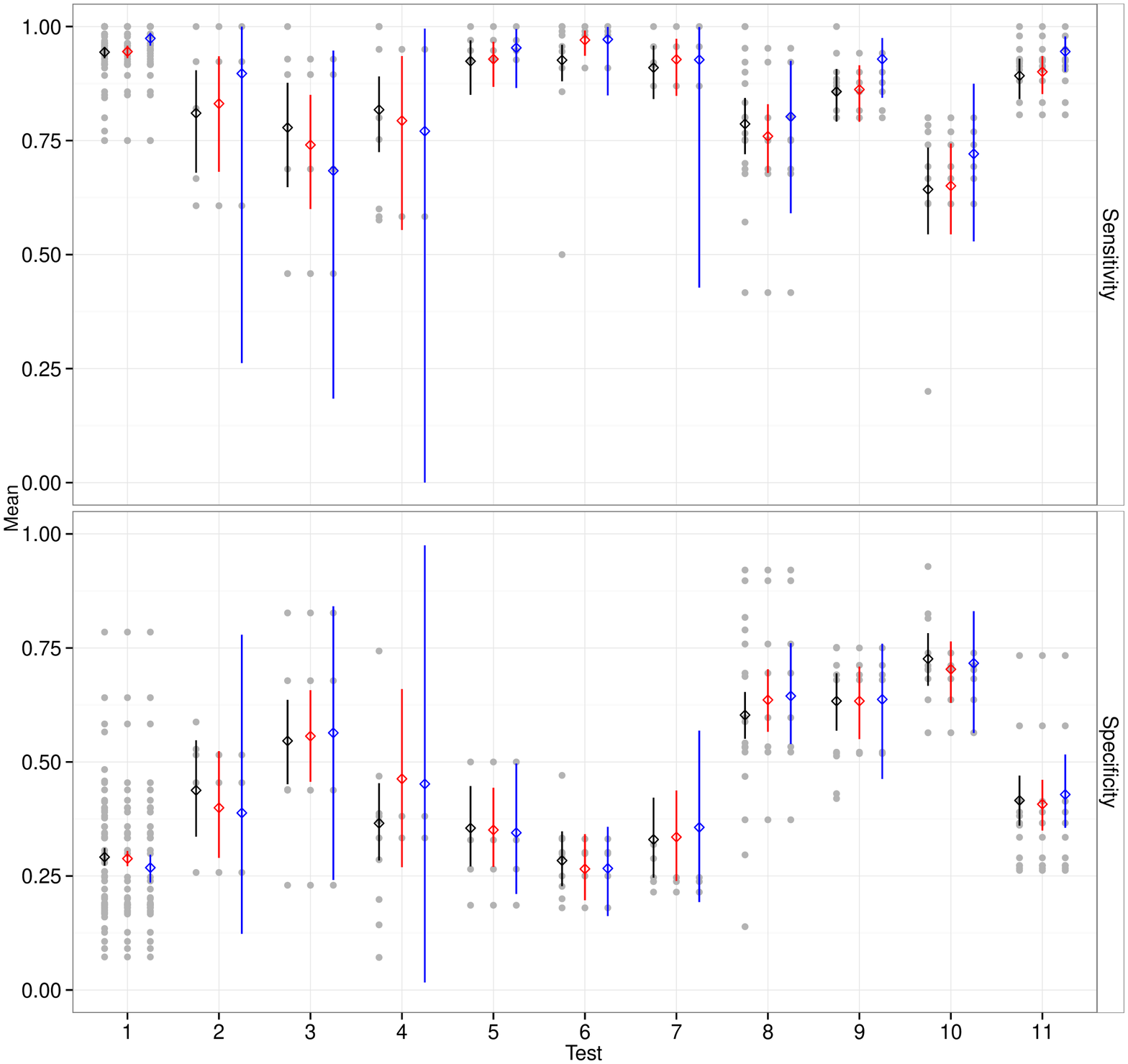}}	
\caption{Plot of study-specific sensitivity (top) and specificity (bottom) in grey points and their corresponding pooled (diamonds) estimates with their 95\% credible intervals (vertical lines) of tests\protect\endnotemark[1] detecting CIN2+\protect\endnotemark[3] in LSIL\protect\endnotemark[2] triage. The black diamonds and vertical lines are estimated by the AB model from all the available studies, the red by the same model but from studies with at least two tests with one of them being test 1 while the blue are estimated by the CB model from studies with at least two tests with one of them being test 1.\label{Fig:4}}
\end{figure}

\begin{table}[h]
	\centering
	\caption{Posterior relative sensitivity and specificity and the corresponding 95\% credible interval of other tests relative to test 1 (HC2) in detecting CIN2+\protect\endnotemark[2] in LSIL\protect\endnotemark[2] triage as estimated by the AB model}
	\label{Tab:3}
	\begin{tabular}{|l|l|l|l|l|l|l|l|}
		\hline
		&                             & \multicolumn{3}{l|}{Relative sensitivity} & \multicolumn{3}{l|}{Relative specificity} \\ \hline
		Label & Index test                  & Mean        & Lower        & Upper        & Mean        & Lower        & Upper        \\ \hline
		2     & Conventional Cytology (CC)  & 0.86        & 0.72         & 0.96         & 1.50        & 1.15         & 1.90         \\ \hline
		3     & Liquid-Based Cytology (LBC) & 0.82        & 0.69         & 0.93         & 1.88        & 1.52         & 2.22         \\ \hline
		4     & Non-Commercial PCR assays   & 0.87        & 0.77         & 0.95         & 1.26        & 0.95         & 1.59         \\ \hline
		5     & Abbott RT PCR hrHPV         & 0.98        & 0.90         & 1.03         & 1.22        & 0.93         & 1.55         \\ \hline
		6     & Linear Array                & 0.98        & 0.93         & 1.02         & 0.98        & 0.77         & 1.21         \\ \hline
		7     & Cobas-4800                  & 0.96        & 0.89         & 1.02         & 1.13        & 0.85         & 1.45         \\ \hline
		8     & P16                         & 0.83        & 0.76         & 0.89         & 2.07        & 1.84         & 2.30         \\ \hline
		9     & P16/Ki67                    & 0.91        & 0.84         & 0.96         & 2.18        & 1.91         & 2.42         \\ \hline
		10    & HPV Proofer(mRNA)           & 0.68        & 0.58         & 0.78         & 2.49        & 2.24         & 2.74         \\ \hline
		11    & APTIMA(mRNA)                & 0.95        & 0.89         & 0.99         & 1.43        & 1.23         & 1.64         \\ \hline
	\end{tabular}
\end{table}

\subsubsection{Triage of women with LSIL to detect CIN3+}
\begin{figure}[h] 
\fbox{\includegraphics[width=\textwidth, height=\textheight,keepaspectratio]{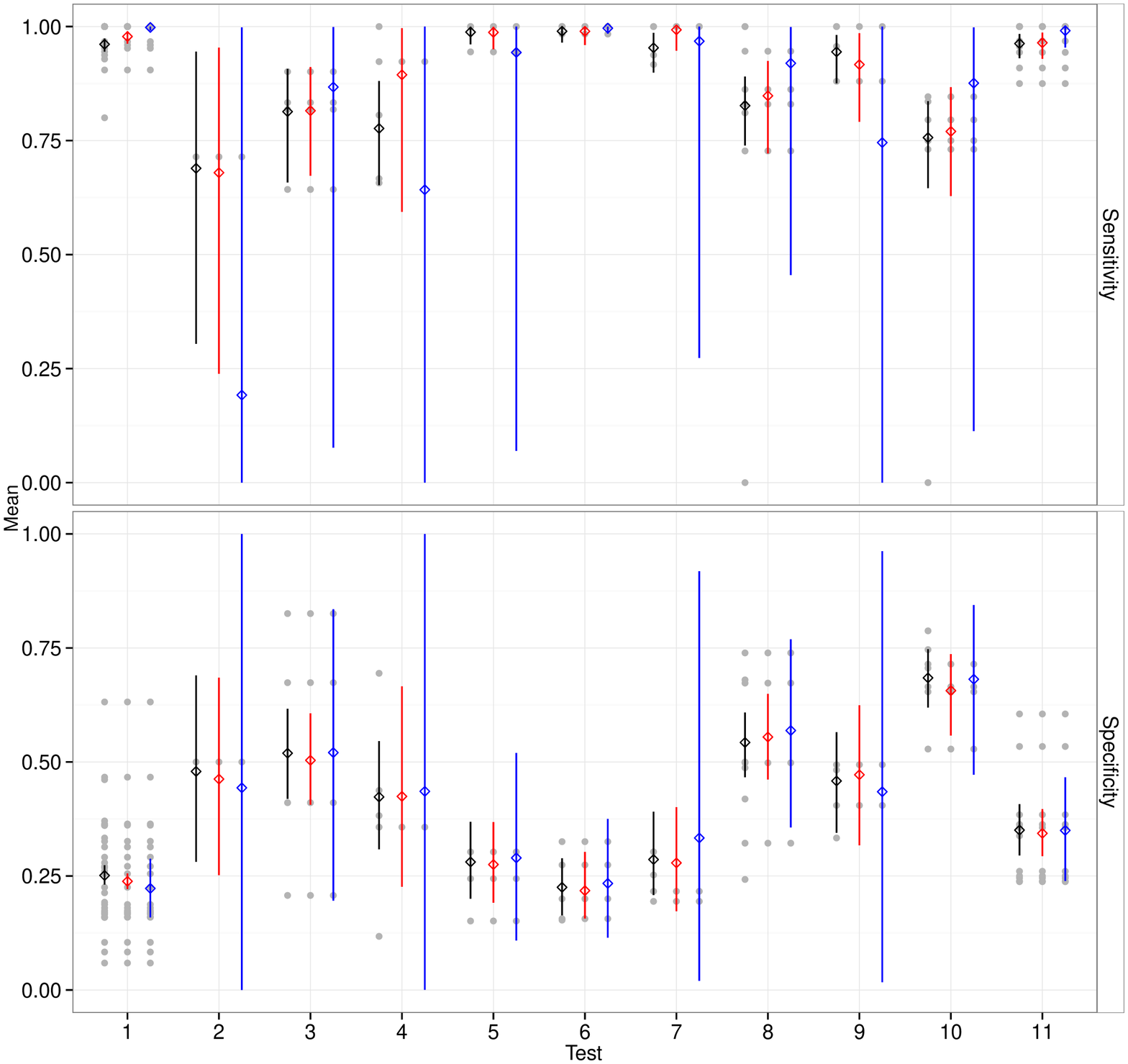}}	
\caption{Plot of study-specific sensitivity and specificity in grey points and their corresponding marginal (black points) sensitivity and specificity with their 95\% credible intervals (vertical lines) of diagnostic tests\protect\endnotemark[1] detecting CIN3+\protect\endnotemark[3] in LSIL\protect\endnotemark[2] triage. The black squares are estimated by the AB model from all the available studies (underlying in grey), the black diamond by the same model but from studies with at least two tests with one of them being test 1 while the black triangles are estimated by the CB model from studies with at least two tests with one of them being test 1.\label{Fig:5}}
\end{figure}

The forest plot presented in figure~\ref{Fig:5} (see also supplementary material:  Results1.xlsx) shows that tests 5 (Abbott RT PCR hrHPV) and 6 (Linear Array) were the most sensitive but among the least specific tests in detecting CIN3+ in women with LSIL. The diagnostic odds ratio indicate that test 5 (Abbott RT PCR hrHPV) had the best discriminatory power (sensitivity = 0.99 [0.96, 1.00], specificity = 0.28 [0.20, 0.37]) while test 9 (p16/Ki67) best discriminatory test (sensitivity = 0.94 [0.88, 0.98], specificity = 0.45 [0.34, 0.56]) according to the superiority index.

According to table~\ref{Tab:3} and \ref{Tab:4} test 5 (Abbott RT PCR hrHPV), 6 (Linear Array) and 7 (Cobas-4800) were as sensitive and as specific while most of the rest of the tests were less sensitive but more specific as test 1 (HC2) in detecting CIN2+ and CIN3+ in triage of women with LSIL cytology.

\begin{table}[h]
	\centering
	\caption{Posterior relative sensitivity and specificity and the corresponding 95\% credible interval of other tests relative to test 1 (HC2) in detecting CIN3+\protect\endnotemark[3] in LSIL\protect\endnotemark[2] triage as estimated by the AB model}
	\label{Tab:4}
	\begin{tabular}{|l|l|l|l|l|l|l|l|l|}
		\hline
		                            & \multicolumn{3}{l|}{Relative sensitivity} & \multicolumn{3}{l|}{Relative specificity} \\ \hline
		Label & Index test                  & Mean        & Lower        & Upper        & Mean        & Lower        & Upper        \\ \hline
		2     & Conventional Cytology (CC)       & 0.71        & 0.31         & 0.99         & 1.91        & 1.13         & 2.76         \\ \hline
		3     & Liquid-Based Cytology (LBC)      & 0.85        & 0.71         & 0.94         & 2.07        & 1.62         & 2.48         \\ \hline
		4     & Non-Commercial PCR assays        & 0.80        & 0.67         & 0.91         & 1.71        & 1.22         & 2.30         \\ \hline
		5     & Abbott RT PCR hrHPV               & 1.03        & 0.99         & 1.05         & 1.11        & 0.79         & 1.50         \\ \hline
		6     & Linear Array                     & 1.03        & 1.00         & 1.05         & 0.90        & 0.63         & 1.19         \\ \hline
		7     & Cobas-4800                       & 0.99        & 0.94         & 1.03         & 1.14        & 0.80         & 1.54         \\ \hline
		8     & P16                               & 0.86        & 0.77         & 0.93         & 2.16        & 1.79         & 2.48         \\ \hline
		9     & P16/Ki67                         & 0.98        & 0.91         & 1.02         & 1.81        & 1.32         & 2.28         \\ \hline
		10    & HPV Proofer(mRNA)               & 0.79        & 0.67         & 0.87         & 2.73        & 2.38         & 3.08         \\ \hline
		11    & APTIMA(mRNA)                    & 1.00        & 0.97         & 1.03         & 1.39        & 1.16         & 1.65         \\ \hline
	\end{tabular}
\end{table}

\subsubsection{Variance Components}
The total variability in sensitivity (in the logit scale) from a compound symmetry working variance-covariance structure ranged from 0.24 [0.04, 0.64] (see supplementary material: Additional-tables.xlsx) in tests used to detect CIN3+ in ASC-US triage to 0.66 [0.41, 1.04] in tests used to detect CIN2+ in LSIL triage. The percentage of total variability in logit sensitivity attributable to between study variability ranged from 21.86\% [0.01\%, 81.19\%] in tests used to detect CIN3+ in LSIL triage to 74.09\% [22.51\%, 99.57\%] in tests used to detect CIN2+ in ASC-US triage.

Similarly for logit specificity, the total variability ranged from 0.39 [0.24, 0.61] in tests used to detect CIN3+ in LSIL to 0.54 [0.40, 0.73] in tests detecting CIN2+ in ASC-US triage. Of the total variability in logit specificitiy, as low as 59.09\% [29.32\%, 79.23\%] in tests used to detect CIN2+ in ASC-US triage and as high as 75.79\% [59.18\%, 88.14\%] in tests used to detect CIN2+ in LSIL triage was due to between study heterogeneity. In other words, there was a stronger correlation between any two logit specificities in a given study than between any two logit sensitivities.

There was in general negative but insignificant correlation between sensitivity and specificity except among tests used to detect CIN2+ in LSIL triage group ($\rho$~ = -0.80 [-1.00, -0.41]). The insignificant correlation parameters suggest absence of overall study effect in the respective data.

\subsubsection{Sensitivity Analysis}
The sensitivity analysis did not highlight any particular change on the mean structure for different priors of the variance-covariance parameters. Based on the MCMC error sampling the variance-covariance $\boldsymbol{\Sigma}$ was better sampled and less auto-correlated with LKJ and Cauchy distributions. 

\subsection{AB versus CB model (black and red vs. blue)}
\begin{figure}[h]
\fbox{\includegraphics[width=\textwidth, height=\textheight,keepaspectratio]{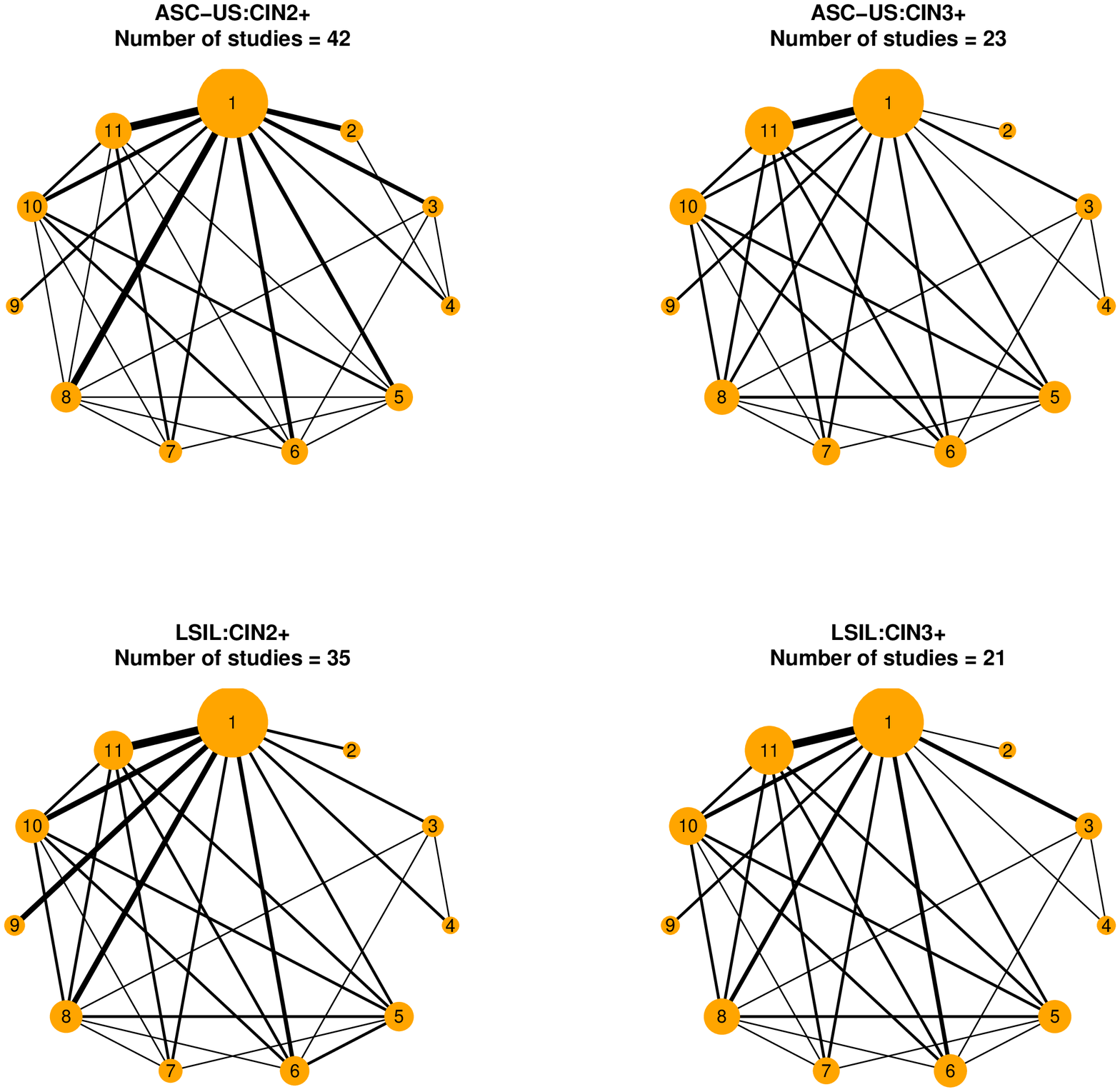}}
\caption{Network plot with studies that evaluated at least two tests\protect\endnotemark[1] with test 1 (HC2) as the common comparator by triage\protect\endnotemark[2] group and outcome\protect\endnotemark[3].\label{Fig:6}}
\end{figure}

The data were re-analysed to compare the estimates from the AB and CB was performed. Studies included in the re-analysis evaluated at least two tests with one of them being test 1 (HC2) set as the common comparator (based on the high number of studies that evaluated test 1 (HC2) besides any another test). Test 1 (HC2) was set as the comparator because most of the studies evaluated it besides another diagnostic test. The network plot of the studies included in the re-analyses is shown in figure~\ref{Fig:6}. 

A graphical summary of the results from the second and third analysis are presented in figures~\ref{Fig:2}, ~\ref{Fig:3}, ~\ref{Fig:4} and ~\ref{Fig:5} and represented by the red and blue diamonds respectively. Overall, there are discrepancies between the locations of the black, red and blue diamonds. 

Firstly, while the black and blue diamonds represent the marginal means, the blue diamonds represent the accuracy estimate for a hypothetical study with random-effects equal to zero. This explains why the black and red diamonds are closer while the blue diamonds are more deviating.

Secondly, the location of the black diamonds is estimated from all available data, including studies evaluating single tests while the location of the red and blue diamonds are determined by studies evaluating at least two studies with one of them being test 1 (HC2). As a consequence of the reduced number of studies, the credible intervals presented in vertical lines are wider especially for the CB approach. As a cascade effect, the ranking of the tests based on the DOR and the superiority also changes (see supplementary material: Results1.xlsx vs. Results2.xlsx vs. Results3.xlsx).

\section{Discussion}
In this paper, we propose a conceptually simple model to estimate sensitivity and specificity of multiple tests within a network meta-analysis framework analogous to a single-factor analysis of variance method with repeated measures. 

The model is based on the assumption that all the tests were hypothetically used but missing at random. When the mechanism of missing data is not a crucial aspect of inference, models ignoring the missing value mechanism and only using the observed data as the proposed model does provide valid answers under a missing at random (MAR) process. In contrast to the CB model, the proposed AB model uses all available data in line with principle of intention-to-treat (ITT) \cite{Fisher}. The missing `unobservable' sensitivities and specificities are parameters are estimated along with the other parameters in the model based on the exchangeability assumption. The cost however is that the model assumptions cannot be formally checked from the data under analysis. 

When the data were never intended to be collected in the first place, the MAR assumption has been shown to hold as is the case in diagnostic studies where older tests become less used and new tests progressively more available with time\cite{Schafer}.

In the analysis, we included studies with at least one test. This is still acceptable because such studies still provide partial information allowing estimation of the mean and the variance-covariance parameters and only the study effects estimates might have larger standard errors~\cite{Gelman07}.
 
The proposed AB model allows for easy estimation of the marginal means and credible intervals for the intra-class correlation. Bayesian methods are known to be computationally intensive but with efficient sampling algorithms such as Hamilton Monte Carlo sampling implemented in Stan~\cite{Stan} convergence to a stationary distribution is accelerated even with poor initial values. Furthermore parallel chain processing greatly reduces computational time.

With the logit transformation, it is assumed that the transformed data is approximately normal with constant variance.  For binary data as well as proportions, the mean and variance depend on the underlying probability. Therefore, any factor affecting the probability will change the mean and the variance. This implies that a linear model where the predictors affect the mean but assume a constant variance will not be adequate. Nonetheless, when the model for the mean is correct but the true distribution is not normal, the maximum likelihood (ML) estimates of the model parameters will be consistent but the standard errors will be incorrect \cite{Agresti}. An alternative to the logit transformation would be a variance stabilizing angular transformation; however the variance stabilizing property of the transform depends on each \textit{n} being large \cite{Crowder}. 

The natural and optimal modelling approach would be to use the beta distribution.  This was the motivation behind our work on copula based bivariate beta distribution in meta-analysis of diagnostic data ~\cite{Nyagaa, Nyagab}. Our further research will focus on how different mean and correlation structures are accommodated and modelled using the beta-binomial distribution in network meta-analysis of diagnostic data. 

There were discrepancies in identifying the best test between the DOR and the superiority index. While the range of values estimated by the two measures range from 0 to infinity, the DOR yield larger values than the superiority index. From the full dataset, the superiority index consistently identified test 9 (p16/Ki67) as the best test. From the reduced data, the DOR identified tests with very low sensitivity but high specificity or vice-versa as the best and in disagreement with the superiority index. This illustrates that DOR cannot distinguish between tests with high sensitivity but low specificity or vice-versa. In contrast, the superiority index gives more weight to tests performing relatively well on both diagnostic accuracy measures and less weight on tests performing poorly on both diagnostic measures or tests performing better on one measure but poorly on the other\cite{Deutsch}. Nonetheless, both measures do not allow to prioritise one parameter which may be clinically appropriate.

Incoherence or inconsistency within NMA is a major concern where for the same contrast, the direct and indirect evidence differ substantially. Lu and Ades (2006) \cite{Lu}, Dias \textit{et al}. (2010) \cite{Dias} and Krahn \textit{et al}.(2014)\cite{Krahn}, explain how to visualize, detect and handle inconsistencies. Since the AB model implicitly assumes consistency, the methods used to detect and quantify inconsistency in CB need not  be used in the AB models.

For the AB models, Hong \textit{et al}. (2015) \cite{Hong} measure inconsistency by data-driven magnitude of bias, the discrepancy between observed and imputed treatment(test) effects while Piepho (2014) \cite{Piepho14} classifies grouping of studies according to the set of tests included and introduce an interaction term: designs by test, to represent inconsistency. We caution against the grouping of studies into designs and including the interaction between designs and test into the model as proposed by Piepho (2014) \cite{Piepho14} because the design variable is an observational factor which will only complicate the `cause-and-effect' inference on test and design.

From our viewpoint, inconsistency/incoherence is a form of heterogeneity between the studies which is often due to missing information in an outlying or influential study. In our model, the influence of the study on the mean is adequately captured by study-effects and the fact that the model hypothetically allows any two tests to be compared directly within each study makes inconsistency less an issue. That said, it is important to assess and identify influence of certain observations on the marginal mean. Detection of influential observations within the Bayesian framework is a computationally involved exercise and still an active research area. 

This article does not consider individual-level data for which the model adaptation is automatic. Future research includes a study on impact of various aspects of data missingness on the robustness of the models.

\section{Conclusion}
The proposed AB model contributes to the knowledge on methods used in systematic reviews of diagnostic data in presence of more than two competing tests. The AB model is more appealing than the CB model for meta-analyses of diagnostic studies because it yields marginal means which are easily interpreted and uses all available data. Furthermore, the model is superior since more general variance-covariance matrix structures can be easily accommodated.

\endnotetext[1]{Tests labels: 1-HC2, 2-CC, 3-LBC, 4-Generic PCR, 5-Abbott RT PCR hrHPV, 6-Linear Array, 7-Cobas-4800, 8-p16, 9-p16/ki67, 10-HPV Proofer, 11-APTIMA.}
\endnotetext[2]{ASC-US: atypical squamous cells of unspecified significance, LSIL :low-grade squamous intraepithelial lesions, CIN: cervival intraepithelial neoplasia}
\endnotetext[3]{CIN2+: cervival intraepithelial neoplasia of grade two or worse, CIN3+: cervival intraepithelial neoplasia of grade three or worse,}
\theendnotes
\bibliographystyle{SageV}
\bibliography{Manuscript}

\begin{funding}
Nyaga V received financial support from the Scientific Institute of Public Health (Brussels) through the OPSADAC project. Arbyn M was supported by the COHEAHR project funded by the 7th Framework Programme of the European Commission (grant No 603019). Aerts M was supported by the IAP research network nr P7/06 of the Belgian Government (Belgian Science Policy). 
\end{funding}

\begin{sm}
		Contact the corresponding author for the supplementary materials.
\begin{enumerate}

\item Model-code.txt  A text file with code of the fitted models in Stan language. 
 
\item Model-fitting.txt   A text file with code to fit the models and reproduce the results.

\item mydata.csv    A file with the data used in this analysis.

  \item Additional-tables.docx A file with additional tables.
  
  \item Results1.xlsx A file with posterior diagnostic accuracy estimates as estimated by the AB model from all the available data.
   
  \item Results2.xlsx A file with posterior diagnostic accuracy estimates as estimated by the AB model from studies that evaluated at least two studies with one of them being test 1.
     
  \item Results3.xlsx A file with posterior diagnostic accuracy estimates as estimated by the CB model from studies that evaluated at least two studies with one of them being test 1.
\end{enumerate}
\end{sm}

\end{document}